\documentclass[twocolumn,showpacs,showkeys,amsmath,amssymb,superscriptaddress,floatfix,nofootinbib]{revtex4}

\usepackage{graphicx}
\usepackage{bm} 
\usepackage{subfigure}

\makeatletter
\newcommand\erfc{\mathop{\operator@font erfc}\nolimits}
\def\slashchar#1{\setbox0=\hbox{$#1$}
   \dimen0=\wd0 \setbox1=\hbox{/} \dimen1=\wd1
   \ifdim\dimen0>\dimen1 \rlap{\hbox to \dimen0{\hfil/\hfil}} #1
   \else  \rlap{\hbox to \dimen1{\hfil$#1$\hfil}} / \fi}

\begin{document}
 
\title{
Initial condition for hydrodynamics, partonic free streaming, and the uniform description of soft observables 
at RHIC%
\footnote{Supported in part by the Polish Ministry of Science and Higher Education, grants N202 153 32/4247 and N202 034 32/0918.}}

\author{Wojciech Broniowski} 
\affiliation{The H. Niewodnicza\'nski Institute of Nuclear Physics, Polish Academy of Sciences, PL-31342 Krak\'ow, Poland}
\affiliation{Institute of Physics, \'Swi\c{e}tokrzyska Academy, ul.~\'Swi\c{e}tokrzyska 15, PL-25406~Kielce, Poland} 

\author{Mikolaj Chojnacki}
\affiliation{The H. Niewodnicza\'nski Institute of Nuclear Physics, Polish Academy of Sciences, PL-31342 Krak\'ow, Poland}

\author{Wojciech Florkowski} 
\affiliation{The H. Niewodnicza\'nski Institute of Nuclear Physics, Polish Academy of Sciences, PL-31342 Krak\'ow, Poland}
\affiliation{Institute of Physics, \'Swi\c{e}tokrzyska Academy, ul.~\'Swi\c{e}tokrzyska 15, PL-25406~Kielce, Poland} 

\author{Adam Kisiel} 
\affiliation{Faculty of Physics, Warsaw University of Technology, PL-00661 Warsaw, Poland}
\affiliation{Department of Physics, Ohio State University, 1040 Physics Research Building, 191 West Woodruff Ave., Columbus, OH 43210, USA}

\date{January 28, 2008}

\begin{abstract}
We investigate the role of the initial condition used for the hydrodynamic evolution of the system formed in ultra-relativistic heavy-ion collisions and find that an appropriate choice motivated by the models of early-stage dynamics, specifically a simple two-dimensional Gaussian profile, leads to a uniform description of soft observables measured in the Relativistic Heavy-Ion Collider (RHIC). In particular, the transverse-momentum spectra, the elliptic-flow, and the Hanbury-Brown--Twiss correlation radii, including the ratio $R_{\rm out}/R_{\rm side}$ as well as the dependence of the radii on the azimuthal angle (azHBT), are properly described. We use the perfect-fluid hydrodynamics with a realistic equation of state based on lattice calculations and the hadronic gas at high and low temperatures, respectively. We also show that the inclusion of the partonic free-streaming in the early stage allows to delay the start of the hydrodynamical description to comfortable times of the order of 1~fm/c. Free streaming broadens the initial energy-density profile, but generates the initial transverse and elliptic flow. The data may be described equally well when the hydrodynamics is started early, or with a delay due to partonic free-streaming. 
\end{abstract}



\maketitle 


The notorious difficulties in simultaneous description of various features of the soft hadron production in the nucleus-nucleus collisions at RHIC, which occur within the standard approach consisting of partonic, hydrodynamic, and hadronic stages, are well known \cite{Heinz:2002un}. In particular, one of the so called RHIC puzzles \cite{Heinz:2002un,Hirano:2004ta,Lisa:2005dd,Huovinen:2006jp} refers to  problems in reconciling the large value of the elliptic flow coefficient, $v_2$, with the Hanbury-Brown--Twiss (HBT) interferometry in numerous approaches including hydrodynamics \cite{Heinz:2001xi,Hirano:2001yi,Hirano:2002hv,Zschiesche:2001dx,Socolowski:2004hw}. In most existing analyses, the large value of $v_2$ prefers long evolution times of the system, while the HBT radii indicate that this time should be short. Typically, the hydrodynamic evolution is initiated from an initial profile generated by Glauber-like models, with the initial temperature serving as a free parameter. 

In this Letter we show that the choice of the initial condition for hydrodynamics is a very important element for the proper description of the data. Ideally, the initial density and flow profiles should be provided by the early partonic dynamics, for instance the Color Glass Condensate (CGC) \cite{McLerran,Kharzeev:2001yq}. In practice, however, the theory of the partonic stage carries some uncertainty in its parameters, moreover, there may exist other effects, see e.g. \cite{Mrowczynski,Muller:2007rs}, which
influence the early dynamics. It is then practical to use some simple parameterization of the initial profile. Here we investigate boost-invariant systems, which approximate well the RHIC collisions at mid-rapidity. We take the following Gaussian parameterization for the initial density profile of the system in the transverse plane $(x_0,y_0)$ at the initial proper time $\tau_0=0.25$~fm,
\begin{eqnarray}
n(x_0,y_0)=\exp \left ( -\frac{x_0^2}{2a^2} -\frac{y_0^2}{2 b^2} \right ). 
\label{profile}
\end{eqnarray} 
The values of the $a$ and $b$ width parameters depend on the centrality and are obtained by matching to the results 
for $\langle x^2 \rangle$ and $\langle y^2 \rangle$ from {\tt GLISSANDO} \cite{Broniowski:2007nz}, which implements the 
{\em shape fluctuations of the system} \cite{Andrade:2006yh}. The values for centrality classes used in this work are collected in Table~\ref{tab:ab}. The profile (\ref{profile}) determines the energy-density profile, from which the temperature profile is obtained \cite{Chojnacki:2007jc}. The initial central temperature, which may depend on the centrality, is denoted by $T_i$ and is a free parameter of our approach. 

\begin{figure}[t]
\begin{center}
\includegraphics[angle=0,width=0.45 \textwidth]{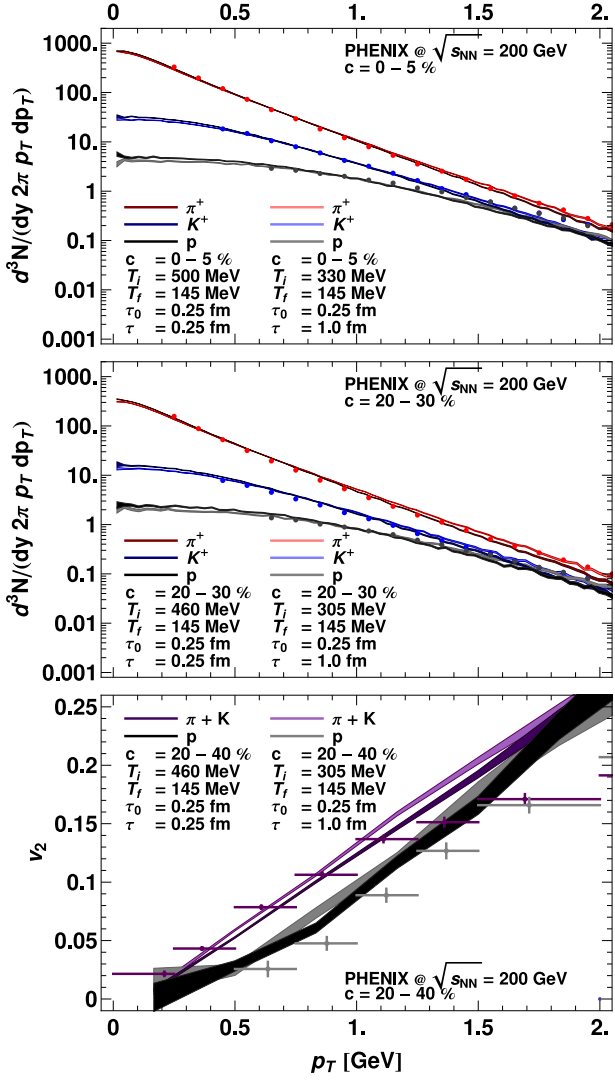}
\end{center}
\vspace{-6.5mm}
\caption{(Color online) The transverse-momentum spectra of pions, kaons and protons for the centrality class $c$=0-5\% (the upper panel) and $c$=20-30\%  (the middle panel), and the transverse-momentum dependence of the elliptic flow coefficient $v_2$ for the centrality class $c$=20-40\% (the lower panel). The darker (lighter) lines/bands describe the model results for the case without (with) free-streaming. The PHENIX data are taken from Refs. \cite{Adler:2003cb,Adler:2003kt}.}
\label{fig:spv2}
\end{figure}

\begin{figure}[t]
\begin{center}
\includegraphics[angle=0,width=0.395 \textwidth]{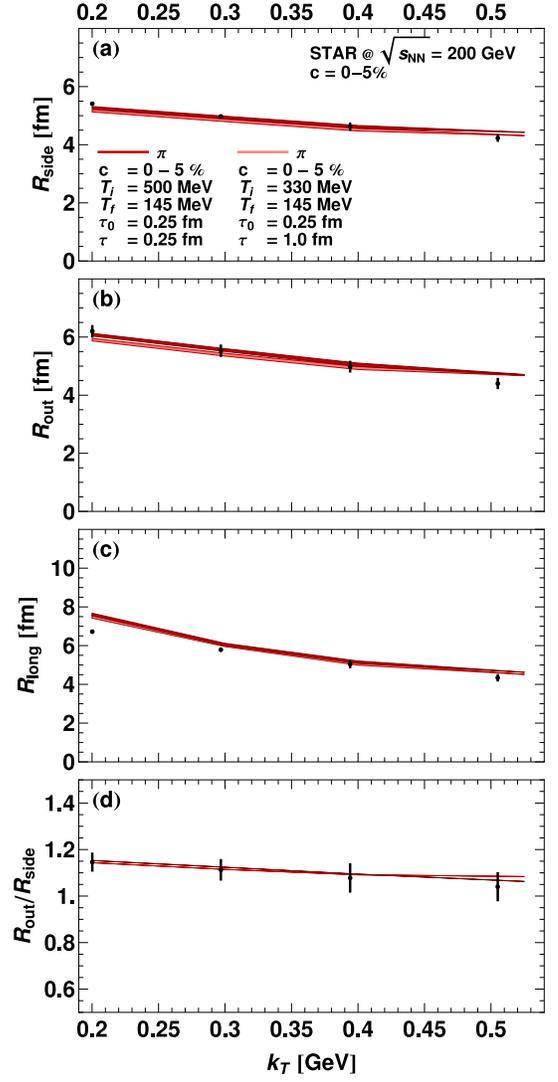}
\end{center}
\vspace{-6.5mm}
\caption{(Color online) The transverse-momentum dependence of the HBT radii $R_{\rm side}$ (a), $R_{\rm out}$ (b), $R_{\rm long}$ (c), and the ratio $R_{\rm out}/R_{\rm side}$ (d)
for central collisions. The darker (lighter) lines describe the results without (with) free-streaming. The STAR data are from Ref. \cite{Adams:2004yc}.}
\label{fig:hbt}
\end{figure}

\begin{table}[b]
\caption{Shape parameters for various centrality classes. \label{tab:ab}}
\begin{tabular}{|r|rrrr|}
\hline
$c$ [\%] & 0-5  & 0-20 & 20-30 & 20-40 \\
$a$ [fm] & 2.65 & 2.41 & 1.94  & 1.78 \\
$b$ [fm] & 2.90 & 2.78 & 2.52  & 2.45 \\
\hline
\end{tabular}
\vspace{-5mm}
\end{table}

The hydrodynamics used is inviscid, baryon-free, and boost-invariant. The equation of state is taken to be as realistic, as possible. We use the lattice QCD simulations of Ref.~\cite{Aoki:2005vt} at high temperatures, $T>170$~MeV, the hadronic gas at $T<170$~MeV, and a smooth interpolation in the vicinity of $170$~MeV, as described in Ref.~\cite{Chojnacki:2007jc}. According to recent knowledge, no first-order transition is implemented, but a smooth cross-over. The hydrodynamic equations are solved by the method of Ref.~\cite{Chojnacki:2006tv}. The accuracy of the method is tested with the entropy conservation, satisfied at the relative level of $10^{-5}$ or better. 
At the temperature $T_f=145$~MeV (model parameter) the system freezes and hadrons (stable and resonances) are generated according to the Cooper-Frye formalism. 
Possible elastic rescattering processes among these hadrons are neglected, thus they stream freely, with the resonances decaying on the way. This stage is simulated with {\tt THERMINATOR} \cite{Kisiel:2005hn}.
With the freeze-out hypersurfaces generated in this work, the collision rate after freeze-out is not very large. For the obtained hypersurfaces the number of pionic trajectory crossings at the distance corresponding to the cross section for the pion collisions is about 1.5-1.7. Hence the single-freeze-out scenario \cite{Broniowski:2001we} seems to be a fairly good approximation for the present case. 
The use of hadronic afterburners for elastic collisions has been described, e.g., in \cite{Teaney:2000cw,Nonaka:2006yn,Hirano:2007xd}.

Our results for central and mid-peripheral collisions with the centrality classes adjusted to the available PHENIX \cite{Adler:2003cb,Adler:2003kt} and STAR \cite{Adams:2004yc} data are shown in Figs.~\ref{fig:spv2} and \ref{fig:hbt} (darker lines/bands). We note a uniform agreement for all soft phenomena studied. In particular, the transverse-momentum spectra, the pionic elliptic-flow, and the HBT radii, including the ratio $R_{\rm out}/R_{\rm side}$, are described within 10\% or better. The HBT results for peripheral collisions are of similar quality 
as in Fig.~\ref{fig:hbt}.

The discussed results have been obtained with an early start of hydrodynamics, at the proper time $\tau=0.25$~fm. It is unlikely that the system should equilibrate so early. Next, we argue that one may delay the starting point of hydrodynamics to realistic times by the inclusion of the partonic free-streaming between the initial proper time $\tau_0=0.25$~fm when the partons are formed and some later time, $\tau$, when hydrodynamics starts. Similar ideas have been described by Sinyukov et al. in Refs.~\cite{Sinyukov:2006dw,Gyulassy:2007zz}. Thus the global picture is as follows: early phase (CGC) generating partons at time $\tau_0$ -- partonic free streaming until $\tau$ -- hydrodynamic evolution until freeze-out at temperature $T_f$ -- free streaming of hadrons and decay of resonances.

Massless partons are formed at the initial proper time  $\tau_0=\sqrt{t_0^2-z_0^2}$ and move along straight lines at the speed of light until the proper time when free streaming ends, $\tau=\sqrt{t^2-z^2}$. We introduce the space-time rapidities \mbox{$\eta_0={1\over2}\log{{t_0-z_0}\over{t_0+z_0}}$} and \mbox{$\eta={1\over2}\log{{t-z}\over{t+z}}$}. Elementary kinematics \cite{Sinyukov:2006dw} links the positions of a parton on the initial and final hypersurfaces and its four-momentum \mbox{$p^\mu=(p_T {\rm cosh} Y, p_T \cos \phi, p_T \sin \phi, p_T {\rm sinh} Y)$}, where $Y$ and $p_T$ are the parton's rapidity and transverse momentum:
\begin{eqnarray}
&&\tau {\rm sinh}(\eta-Y)=\tau_0 {\rm sinh}(\eta_0-Y),  \label{kinem} \\
&&x=x_0+\Delta  \cos \phi, \;\;  y=y_0 + \Delta \sin \phi  , \nonumber \\
&& \Delta =\frac{t-t_0}{{\rm cosh}Y}=\tau {\rm cosh}(Y-\eta)-\sqrt{\tau_0^2+\tau^2 {\rm sinh}^2(Y-\eta)}. \nonumber
\end{eqnarray} 
Thus the phase-space density of partons at the proper times $\tau_0$ and $\tau$ are related,
\begin{eqnarray}
&&\frac{d^6N(\tau)}{dY d^2p_T d\eta dx dy} = \int d \eta_0 dx_0 dy_0 \frac{d^6N(\tau_0)}{dY d^2p_T d\eta_0 dx_0 dy_0} \times \nonumber \\ 
&& \delta(\eta_0-Y-{\rm arcsinh} [\frac{\tau}{\tau_0} {\rm sinh}(\eta-Y)] ) \times \label{fs} \\
&&\delta(x-x_0-\Delta  \cos \phi)\delta( y-y_0- \Delta \sin \phi). \nonumber
\end{eqnarray}  
It is reasonable to assume a factorized boost-invariant form of the initial distribution of partons, 
\begin{eqnarray}
 \frac{d^6N(\tau_0)}{dY d^2p_T d\eta_0 dx_0 dy_0} = n(x_0,y_0) F(Y-\eta_0,p_T),
\end{eqnarray}
where $n$ is the transverse density of Eq.~(\ref{profile}). When the emission profile $F$ is focused near $Y=\eta_0$, for instance \mbox{$F \sim \exp[-(Y-\eta_0)^2/(2 \sigma^2)]$}, with $\sigma \sim 1$, and if $\tau \gg \tau_0$, then  the kinematic condition (\ref{kinem}) 
transforms it into 
\mbox{$F \sim \exp \left ( -{{\rm arcsinh}^2 \left [ \frac{\tau}{\tau_0} \sin(Y-\eta) \right ]}/({2 \sigma^2}) \right )$}
which is so sharply peaked that effectively $F \sim \delta(Y-\eta)$.
Then Eq.~(\ref{fs}) yields
\begin{eqnarray}
\frac{d^6N(\tau)}{dY d^2p_T d\eta dx dy} &=& n(x-\Delta \tau \cos \phi,y-\Delta \tau \sin \phi) \times \nonumber \\
&&\delta(Y-\eta)f(p_T). 
\end{eqnarray}
where $\Delta \tau=\tau -\tau_0$ and $f(p_T)$ is the transverse momentum distribution. 
\begin{figure}[tb]
\begin{center}
\includegraphics[angle=0,width=0.37 \textwidth]{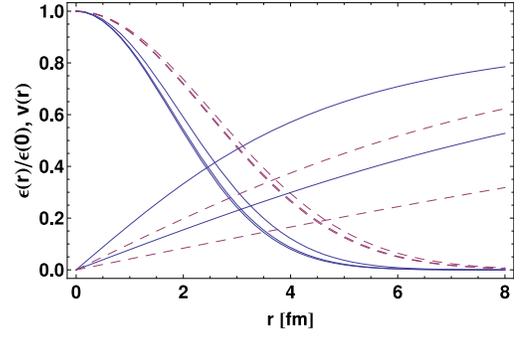}
\end{center}
\vspace{-6.5mm}
\caption{(Color online) Sections of the energy-density profile $\epsilon$ (gaussian-like curves) normalized to unity at the origin, and of the velocity profile $v=\sqrt{v^2_x+v^2_y}$ (curves starting at the origin), cut along the $x$ axis (solid lines) and $y$-axis (dashed lines). The initial profile is from Eq.~(\ref{profile}) for centrality 20-40\% at $\tau_0=025$~fm. The $\epsilon$ profiles are for $\tau=\tau_0=0.25$, $1$, and $2$~fm,  while the velocity profiles are for $\tau=1$ and $2$~fm, all from bottom to top. We note that the flow is azimuthally asymmetric and stronger along the $x$ axis.
\label{fig:fs}}
\end{figure}
The energy-momentum tensor at the proper time $\tau$, rapidity $\eta$, and transverse position $(x,y)$ is given by the formula
\begin{eqnarray}
&&T^{\mu \nu}=\int dY d^2p_T  \frac{  d^6N(\tau)}{dY d^2p_T d\eta dx dy} p^\mu p^\nu  \label{tmunu} \\
&&=A \int_0^{2 \pi} d\phi \, n\left(x-\Delta \tau \cos \phi,y-\Delta \tau \sin \phi\right) \times \nonumber \\
&&\left ( \begin{array}{cccc} 
{\rm cosh}^2 \eta &  {\rm cosh}\eta \cos \phi &  {\rm cosh}\eta \sin \phi & {\rm cosh}\eta {\rm sinh}\eta\\ 
{\rm cosh}\eta \cos \phi & \cos^2 \phi & \cos \phi \sin \phi & \cos \phi {\rm sinh} \eta \\ 
{\rm cosh} \eta \sin \phi & \cos \phi \sin \phi & \sin^2 \phi & \sin \phi {\rm sinh} \eta \\
{\rm cosh}\eta {\rm sinh}\eta & \cos \phi {\rm sinh}\eta & \sin \phi {\rm sinh}\eta & {\rm sinh}^2\eta \end{array} \right ), \nonumber 
\end{eqnarray}
where $A$ is a constant from the $p_T$ integration. Due to boost invariance the further calculation may be carried 
at $\eta=0$. 
Next, we assume that at the proper time $\tau$ the system {\em equilibrates rapidly}. We thus use the {\em Landau matching condition},  
\begin{eqnarray}
T^{\mu \nu}(x,y) u_\nu(x,y) =\epsilon(x,y) g^{\mu \nu} u_\nu(x,y), \label{landau}
\end{eqnarray}
which states that at each point the four-velocity of the fluid, $u^\mu=(1,v_x,v_y,0)/\sqrt{1-v^2}$, is such that $T^{\mu \nu}$ is diagonal in the local rest frame of the matter. The position-dependent eigenvalue $\epsilon$ is identified with the energy-density profile. The result of solving Eq.~(\ref{landau}) with $T^{\mu \nu}$ from (\ref{tmunu}) for $\Delta \tau =0.75$~fm and the initial profile (\ref{profile}) for $c$=20-40\% is shown in Fig.~\ref{fig:fs}. The curves  
show the sections along the $x$ and $y$ axes of $\epsilon$ and the velocity $v$ at $\tau_0$ (no free streaming) and $\tau=1$ and 2~fm. Obviously, at $\tau_0$ we find $\epsilon(x,y) = 2\pi A n(x,y)$. We note that, naturally, the profile spreads out as the time progresses. Importantly, this effect is faster along the shorter axes, $x$. This is clearly indicated by the velocity profiles along the $x$ and $y$ axes. Thus the {\em flow generated by free streaming is azimuthally asymmetric}. It can also be obtained upon the low $\Delta \tau$ and low $x,y$ expansion, where straightforward algebra gives (for $x \Delta \tau \ll a^2$ and $y\Delta \tau \ll b^2$)
\begin{eqnarray}
{\bf v}(x,y)=-\frac{\Delta \tau}{3} \frac{\nabla n(x,y)}{n(x,y)} 
&=&\frac{\Delta \tau}{3} \left ( \frac{x}{a^2}, \frac{y}{b^2},0 \right ).
\end{eqnarray}

The results of the hydrodynamic calculation starting from free-streaming down to $\tau=1$~fm and readjusted initial temperature at 
the proper time $\tau$, followed 
by {\tt THERMINATOR} simulations, are shown in Figs.~\ref{fig:spv2} and \ref{fig:hbt} with lighter curves/bands. We notice very similar 
results to the case with no free streaming, and again a proper description of the data. Larger free-streaming times spoil this agreement, as the flow becomes too strong. We remark, however, that the departure from the $\delta(Y-\eta)$ condition weakens the flow, which allows for larger values of $\tau$.
The basic conclusion here is that the partonic free streaming may be used to {\em delay the start of hydrodynamics}. The physical results are basically unaltered, as the dispersion of the density profile, resulting in milder hydrodynamic development of flow, is accompanied by the 
buildup of the {\em initial flow}.

\begin{figure}[tb]
\begin{center}
\includegraphics[angle=0,width=0.44 \textwidth]{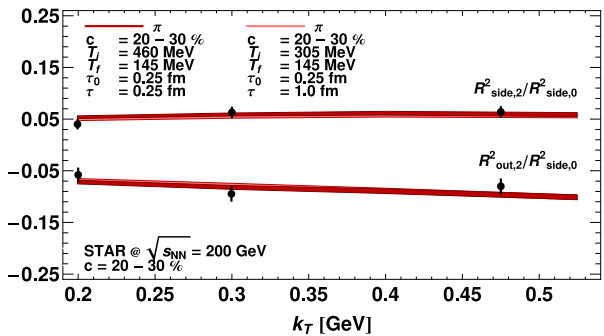}
\end{center}
\vspace{-6.5mm}
\caption{(Color online) Study of azHBT: The quantities $R^2_{{\rm side},2}/R^2_{{\rm side},0}$ and $R^2_{{\rm out},2}/R^2_{{\rm side},0}$ from the model (bands) and experiment \cite{Adams:2003ra} (points), plotted as functions of the transverse momentum of the pion pair.
\label{fig:azHBT}}
\end{figure}
Having properly described the HBT radii, we may now 
deal ``as a bonus'' with the azimuthally sensitive HBT interferometry \cite{Adams:2003ra} and consider 
the averages over the azimuthal angle, $R^2_{i,2}(k_T) = \langle R^2_{i}(k_T,\varphi) \cos(2\varphi) \rangle$, where $i=$ side or out 
($R_{2,{\rm long}}=0$). The results shown in Fig.~\ref{fig:azHBT} display a remarkable agreement between our model and the data (the curves with 
and without free streaming overlap).

In conclusion: 1) With a proper choice of the initial profile (e.g. Gaussian) one may obtain a uniform description of soft observables ($p_T$-spectra, $v_2$, and pionic HBT radii) at RHIC with the boost-invariant inviscid hydrodynamics with a realistic equation of state and thus solve one of the ``RHIC puzzles''. 2)~Partonic free streaming 
generates initial transverse and elliptic flow. It may be used to delay the start of the hydrodynamic phase. 3)~Azimuthally-sensitive HBT 
is described in agreement with the data. 4)~The complete treatment of resonances is important, as is well known from statistical models. 
5)~When other effects (viscosity, departure from boost invariance) are incorporated, the important role of the initial condition should not be overlooked.

\end{document}